\documentclass[sigconf, screen]{acmart}

\AtBeginDocument{%
  \providecommand\BibTeX{{%
    \normalfont B\kern-0.5em{\scshape i\kern-0.25em b}\kern-0.8em\TeX}}}

\setcopyright{acmcopyright}
\copyrightyear{2019}
\acmYear{2019}
\acmDOI{xx.xxx/xxxxx.xxxxxxx}

\acmBooktitle{ACM Trans. Knowl. Discov., June 03--05, 2019}
\acmPrice{15.00}
\acmISBN{xxx-x-xxxx-xxx-x/xx/xx}

\usepackage{graphicx}
\usepackage{subfigure}

\begin{document}

\title{Attention: to Better Stand on the Shoulders of Giants}


\author{Sha Yuan}
\orcid{0000-0002-5503-3008}
\affiliation{%
  \institution{Department of Computer Science and Technology, Tsinghua University}
  \streetaddress{30 Shuangqing Rd}
  \city{Haidian Qu}
  \state{Beijing Shi}
  \country{China}
}
\email{yuansha@mail.tsinghua.edu.cn}

\author{Zhou Shao}
\affiliation{%
  \institution{School of Computer Science and Engineering, Nanjing University of Science and Technology}
}
\email{shaozhou@mail.tsinghua.edu.cn}

\author{Yu Zhang}
\affiliation{%
  \institution{Institute of Medical Information, Peking Union Medical College, Chinese Academy of Medical Sciences}
}
\email{zhang.yu@imicams.ac.cn}

\author{Xingxing Wei}
\affiliation{%
 \institution{Department of Computer Science and Technology, Tsinghua University}
 \streetaddress{30 Shuangqing Rd}
  \city{Haidian Qu}
  \state{Beijing Shi}
  \country{China}
}
\email{xwei11@mail.tsinghua.edu.cn}

\author{Tong Xiao}
\affiliation{%
  \institution{Department of Computer Science and Technology, Tsinghua University}
  \streetaddress{30 Shuangqing Rd}
  \city{Haidian Qu}
  \state{Beijing Shi}
  \country{China}}
\email{abravesailor@gmail.com}

\author{Yifan Wang}
\affiliation{%
  \institution{Department of Computer Science and Technology, Tsinghua University}
  \streetaddress{30 Shuangqing Rd}
  \city{Haidian Qu}
  \state{Beijing Shi}
  \country{China}}
\email{yifan-wa16@mails.tsinghua.edu.cn}

\author{Jie Tang}
\authornote{Jie Tang is the corresponding author.}
\affiliation{
  \institution{Department of Computer Science and Technology, Tsinghua University}
  \streetaddress{30 Shuangqing Rd}
  \city{Haidian Qu}
  \state{Beijing Shi}
  \country{China}}
\email{jietang@tsinghua.edu.cn}

\renewcommand{\shortauthors}{Sha and Jie, et al.}

\begin{abstract}
Science of science (SciSci) is an emerging discipline wherein science is used to study the structure and evolution of science itself using large data sets. The increasing availability of digital data on scholarly outcomes offers unprecedented opportunities to explore SciSci. 
In the progress of science, the previously discovered knowledge principally inspires new scientific ideas, and citation is a reasonably good reflection of this cumulative nature of scientific research.
The researches that choose potentially influential references will have a lead over the emerging publications. 
Although the peer review process is the mainly reliable way of predicting a paper's future impact, the ability to foresee the lasting impact based on citation records is increasingly essential in the scientific impact analysis in the era of big data. This paper develops an attention mechanism for the long-term scientific impact prediction and validates the method based on a real large-scale citation data set. The results break conventional thinking. Instead of accurately simulating the original power-law distribution, emphasizing the limited attention can better stand on the shoulders of giants.
\end{abstract}



\maketitle

\section{Introduction}

\begin{figure}[htb] 
\centering
\subfigure[The volume of literatures.]{
\includegraphics[width=0.48\textwidth]{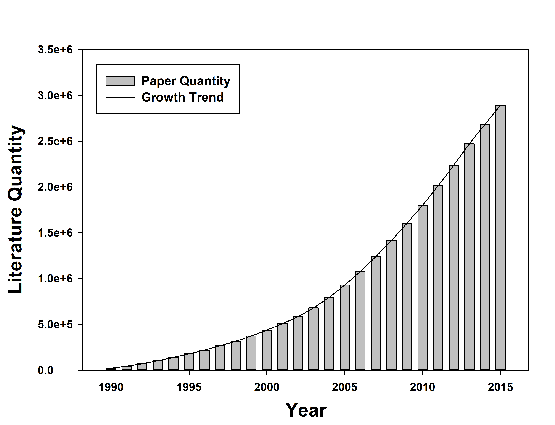}
\label{fig_paperStatistic}}
\subfigure[The citation Distribution.]{
\includegraphics[width=0.48\textwidth]{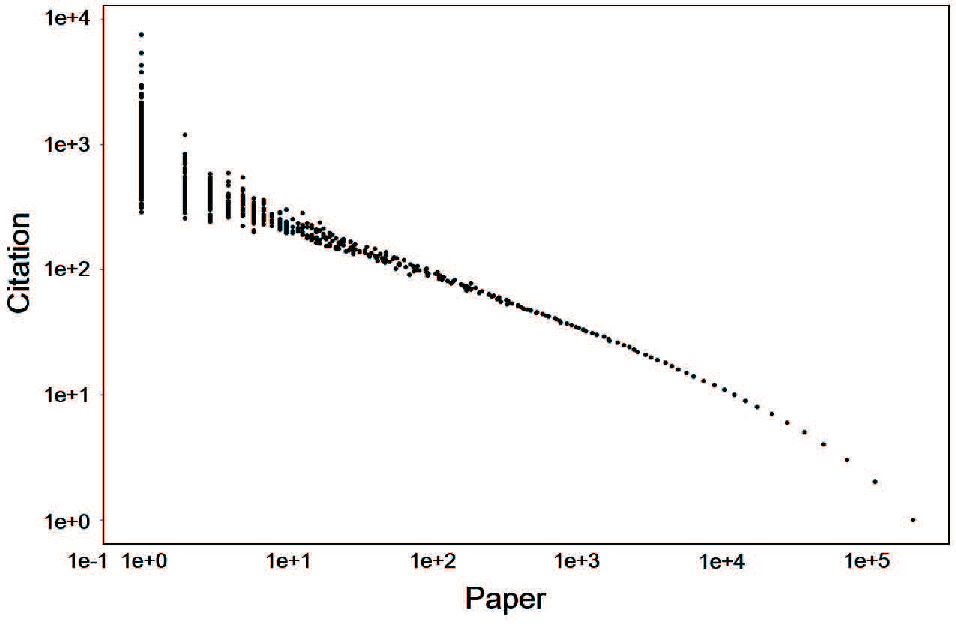}
\label{fig_citationStatistic}}
\caption{Statistics of literature data from AMiner.}
\label{fig_paperStatisticPlot}
\end{figure}

The increasing availability of digital data on scholarly outcomes offers unprecedented opportunities to explore the science of science (SciSci) \cite{fortunato2018science}. 
Based on the empirical analysis of big data, SciSci provides a quantitative understanding of scientific discovery, creativity, and practice. It discovers that the previously discovered knowledge mainly inspires new scientific ideas, and citation is a relatively good reflection of this cumulative nature of scientific research. Citation count, which has been used to evaluate the quality and influence of scientific work for a long time, stands out from many quantification measure metrics of scientific impact.
With the rapid evolution of scientific research, there is a huge volume of literature published every year, and this situation is expected to remain within the foreseeable future. Fig.~\ref{fig_paperStatisticPlot} shows the statistics on AMiner~\cite{Tang2008ArnetMiner}, which is a large literature database in Computer Science. Fig.~\ref{fig_paperStatistic} visualizes the explosive increase on the volume of publications in the past years from $1990$ to $2015$. It shows that the literature quantity assumes the exponential order to grow.

Useful scientific research requires reviewing the previous researches. It is not wise, nor possible, for researchers to track all existing related work due to the enormous volume of the existing publications. In general, researchers follow or cite merely a small proportion of high-quality publications. SciSci provides several quantification methods for scientific impact measurement in article-level, author-level, and journal-level. Much SciSci work has been done on the evaluation metrics for the quality and influence of scientific work, including citation count, h-index \cite{Hirsch2005An}, and impact factor \cite{garfield2001impact}. One of the most basic quantification measure metrics of scientific impact is citation count. It measures the number of received citations for an article. Many other essential evaluation criteria of authors (e.g., h-index) and journals (e.g., Impact Factor) are calculated based on citation count.

A lot of SciSci researchers have focused on the characterization of scientific impact, such as the universal citation distributions \cite{radicchi2008universality}, the characteristics of citation networks \cite{hunter2011dynamic,pan2012world,kuhn2014inheritance}, and the growth pattern of scientific impact \cite{dong2015will}. The results reveal the regularity of scientific progress that a few research papers attract the vast majority of citations \cite{barabasi2012publishing}, long-distance interdisciplinarity leads to higher scientific impact \cite{lariviere2015long,yegros2015does}.
Fig.~\ref{fig_citationStatistic} illustrates the citation distribution (the number of papers vs. citation counts) of about two million papers in AMiner. It is natural to find that not all publications attract equal attention in academia. A few research papers accumulate the vast majority of citations, and most of the other papers attract only a few citations~\cite{barabasi2012publishing}.
The citation distribution follows the power-law distribution. A small number of scholarly outcomes are more likely to attract scientists' attention than the others accounting for a vast majority. For the ever-growing literature quantity, it is significative to forecast which paper is more likely to attract more attention in academia.

The fact is that the current citation count and the derived metrics can only capture the past accomplishment. They lack the predictive power to quantify future impact~\cite{acuna2012future}. Predicting an individual paper's citation count over time is significant, but (arguably) very difficult. To predict the citation count of individual items within a complex evolving system, current models are falling into two main paradigms. One formulates the citation count over time as time series, and then makes predictions by either exploiting temporal correlations~\cite{Szabo2010Predicting}, or fitting these time series with certain classes of designed functions~\cite{Matsubara2012Rise,Bao2013Popularity}, including the regression models~\cite{Yan2011Citation}, the counting process~\cite{Vu2011Dynamic}, the point process, the Poisson process~\cite{wang2013quantifying}, Reinforced Poisson Process (RPP)~\cite{shen2014modeling}, self-excited Hawkes Process~\cite{Bao2016Modeling}, RPP with self-excited Hawkes Process~\cite{Xiao2016}. The designed functions consider various factors.

The other prevalent line utilize Deep Neural Network (DNN) based models to solve the scientific impact prediction problem. Recently, Convolutional Neural Network (CNN) and Recurrent Neural Network (RNN), have received considerable attention from both the academia and the industry. RNN has been proven to perform particularly well on temporal data series~\cite{sutskever2014sequence}. Due to the vanishing gradient problem, RNN always fails to handle the temporal contingencies present in the input/output sequences spanning long intervals~\cite{bengio1994learning}. The networks with loops in RNN allow information to persist in a long time.
Long short-term memory (LSTM) is proven to be capable of learning long-term dependencies. RNN with LSTM units performs rather well in handling long-term temporal data series~\cite{Xiao2017Modeling}.

All the existing methods try to tune the citation distribution exactly as the power-law distribution. However, this paper argues that the effectiveness of quantifying long-term scientific impact is fundamentally limited in this line of thinking. This paper proposes to put more attention on some specific items. The authors validate the proposed line of thinking on a real large-scale citation dataset. Extensive experiment results demonstrate that the proposed method possesses remarkable power at predicting the long-term scientific citation. The most important contribution is that this paper is the first to change the line of thinking in quantifying the long-term scientific impact. Instead of simulating the power-law distribution, researchers need to pay more attention to the limited attention to better stand on the shoulders of giants.

\section{Problem Formulation}
The basic evaluation metric for scientific impact is citation count. The received citation count of an individual paper $d$ during time period $[0, T]$ is characterized by a time-stamped sequence $\{ n_{d}^{t} \}_{t=0}^{T}$, where $n_{d}^{t}$ represents the number of citation counts received by paper $d$ at time $t$, $n_{d}^{t}$ is an integer greater than or equal to zero. In the context of giving the historical citation records, the goal is to model the future citation count and predict it over an arbitrary time.

\begin{definition}
    \textbf{Scientific Impact.} Given the literature corpus $D$, $\mathrm{card}(D)$ $=M$, the scientific impact of a literature article $d\in D$ at time $t$ is defined as its citation counts $n_{d}^{t}$:
    \begin{equation}
    \begin{aligned}
    citing_d^t &=  \{\tilde{d} \in D, \tilde{d}\neq d:\tilde{d}^t~cites~d\},\\
    n_{d}^{t} &= \mathrm{card}(citing_d^t).
    \end{aligned}
    \end{equation}
\end{definition}

The underlying assumption of the citation count here is the accumulated citations, which make it possible to quantify citations for different items at different times. The long-term scientific impact of individual item $d$ can be formalized as the following time series $\{n_{d}^{0},\cdots , n_{d}^{t},\cdots, n_{d}^{T}\}$.
Without loss of generality, the number of accumulated citation count increase over time. And then, we have $0=n_{d}^{0}\leq \cdots \leq n_{d}^{t}\leq \cdots \leq n_{d}^{T}=N_d$.

The \textbf{scientific impact prediction problem} can be formalized as follows.

\textbf{Input:} For each paper $d$, the input is $\{(x_{d}^{0},n_{d}^{0}),\cdots ,$ $(x_{d}^{t},n_{d}^{t}),\cdots\}$ $\in \mathbb{N}^K \times \mathbb{N}$, where $\vec{X}=\{x_{d}^{0},\cdots ,x_{d}^{t},\cdots\}$, and $x_{d}^{t}$ is expressed as a $K$-dimensional feature vector, and $n_{d}^{t}$ is the citation counts of paper $d$ at time $t$.

\textbf{Learning:} The goal of citation count prediction is to learn a predictive function $f$ ($\mathbb{N}^K \rightarrow \mathbb{N}$) to predict the citation counts of an article $d$ after a given time period $t$. Formally, we have
\begin{equation}
f(d|\vec{X}, t)\rightarrow \hat{n}_d^{t},
\end{equation}
where $\hat{n}_d^{t}$ is the predicted citation count and $n_{d}^{t}$ is the actual one.

\textbf{Prediction:} Based on the learned prediction function, we can predict the citation count of a paper for the next years, for example, the citation count of paper $d$ at time $t$ is given by $f(d|\vec{X}, t)$.

\begin{figure*}[htb] 
\centering
\subfigure[The overview.]{
\includegraphics[width=0.3\textwidth]{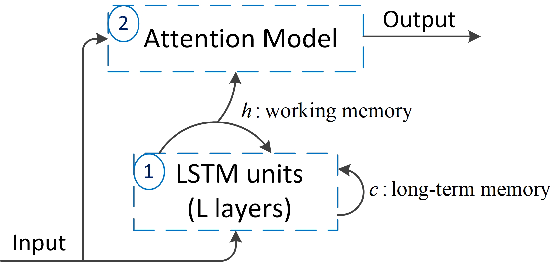}
\label{subfig_overview}}
\subfigure[RNN with LSTM units.]{
\includegraphics[width=0.3\textwidth]{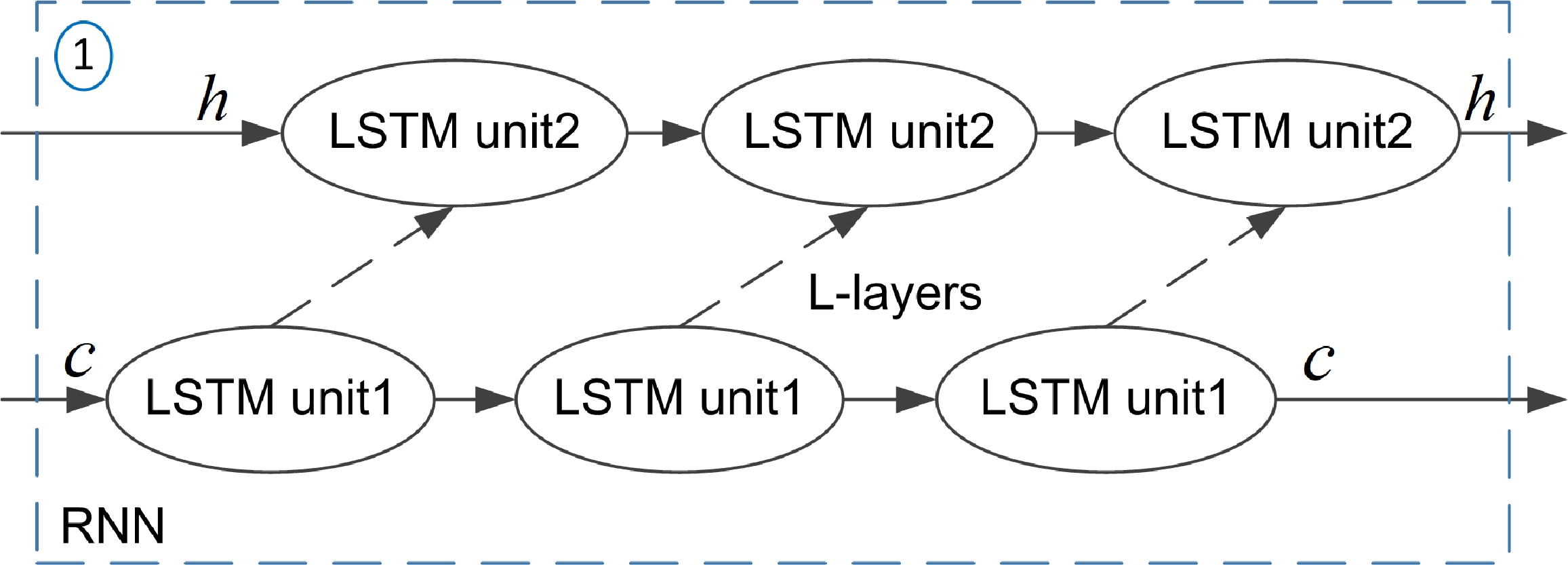}
\label{subfig_LSTM_att}}
\subfigure[The attention model.]{
\includegraphics[width=0.28\textwidth]{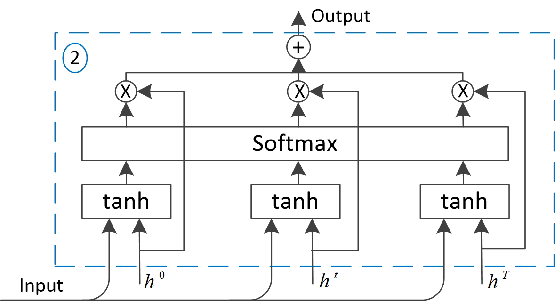}
\label{subfig_attention}}
\caption{The deep learning attention model.}
\label{fig_model2}
\end{figure*}

\section{Scientific Impact Prediction}\label{sec:model}
As the most efficient scientific impact prediction method found so far, RNN has already achieved compelling performance in predicting the scientific impact. This paper embeds the RNN with LSTM units as a baseline and then emphasize highly cited papers in the proposed attention mechanism. Although many other fields have used the attention mechanism, the proposed method gives the new insight about long-term quantifying scientific impact. Instead of adapting citation distribution to a power-law distribution, the findings in this paper provide a new line of thinking for the SciSci research.

\subsection{Deep Learning Attention Mechanism}
Given a time-stamped sequence $\{ n_{d}^{t} \}_{t=0}^{T}$, a $K$-dimensional feature vector $\vec{X}=\{x_{d}^{0},\cdots ,x_{d}^{t},\cdots ,x_{d}^{T}\}$ needs to be designed as input.
The input space of every item with popularity records $\{(x^{0},n^{0}),$ $\cdots ,(x^{t},n^{t}),\cdots ,(x^{T},n^{T})\}$ reflects the intrinsic quality of the item. Fig.~\ref{subfig_overview} gives an overview of the model architecture. There are two key components in the architecture: the RNN with LSTM units and the attention model. As illustrated in Fig.~\ref{subfig_LSTM_att}, it arranges the LSTM units in the form of RNN with $L$ layers. In the deep neural network, the parameter $L$ depends on the input scale. RNN is famous for its popularity and well-known capability for efficient time series learning~\cite{Xiao2017Modeling}.
The LSTM units capture the long-range dependency in long-term scientific impact quantification.

\vspace{2ex} 
\textbf{The RNN with LSTM Units}. The LSTM units are arranged in the form of RNN, as illustrated in Fig.~\ref{subfig_LSTM_att}. There are four major components in a standard LSTM unit, including a memory cell, a forget gate $\Gamma_f$, an input gate $\Gamma_i$, and an output gate $\Gamma_o$. The gates are responsible for information processing and storage over arbitrary time intervals.
Usually, the outputs of these gates are between $0$ and $1$. A new study gives suggestions to push the output values of the gates towards $0$ or $1$. By doing so, the gates are mostly open or closed, instead of in a middle state~\cite{li2018towards}.
This paper arranges the LSTM units in the form of RNN. In this way, introducing the memory cell will solve the vanishing gradient problem. Thus, it can store information for either short or long periods in the LSTM unit.

Intuitively, the input gate controls the extent to which a new value flows into the memory cell. A function of the inputs passes through the input gate and is added to the cell state to update it.
The following formula for the input gate is used:
\begin{equation}
\Gamma _{i}^{t}=\sigma \left ( W_i\left [ h^{t-1},x^{t} \right ] +b_{i}\right ),
\end{equation}
where matrix $W_i$ collects the weights of the input and recurrent connections. The symbol $\sigma$ represents the Sigmoid function. The values of the vector $\Gamma _{i}^{t}$ are between $0$ and $1$. If one of the values of $\Gamma _{i}^{t}$ is $0$ (or close to $0$), it means that this input gate is closed and no new information is allowed into the memory cell at time $t$. If one of the values is $1$, the input gate is open for new coming value at time $t$. Otherwise, the gate is in the state of half-open half-clearance.

The forget gate controls the extent to which a value remains in the memory cell. It provides a way to get rid of the previously stored memory value. Here is the formulation of the forget gate:
\begin{equation}\label{eq_forget}
\Gamma _{f}^{t}=\sigma \left ( W_f\left [ h^{t-1},x^{t} \right ] +b_{f}\right ),
\end{equation}
where $W_f$ is the weight matrix that governs the behavior of the forget gate. Similar to $\Gamma _{i}^{t}$, $\Gamma _{f}^{t}$ is also a vector of values between $0$ and $1$. If one of the values of $\Gamma _{f}^{t}$ is $0$ (or close to $0$), it means that the memory cell should remove that piece of information in the corresponding component in the cell. If one of the values is $1$, the corresponding information will be kept.

Remembering information for long periods of time is practically the default behavior of LSTM. The long-term accumulative influence is formulated as follows:
\begin{equation}\label{eq_long}
c^{t}=\Gamma _{f}^{t}\ast c^{t-1}+\Gamma _{i}^{t}\ast\tilde{c}^{t},
\end{equation}
where $\ast$ denotes the Hadamard product (the element-wise multiplication of matrices), $\tilde{c}^{t}$ is calculated as follows:
\begin{equation}
\tilde{c}^{t}=\tanh\left ( W_c\left [ h^{t-1},x^{t} \right ] +b_{c}\right ).
\end{equation}
That is, the information in memory cell consists of two parts: the retained old information $\Gamma _{f}^{t}\ast c^{t-1}$ (controlled by the forget gate), and the new coming information $\Gamma _{i}^{t}\ast\tilde{c}^{t}$ (controlled by the input gate).

The output gate controls the extent to which the value in the cell is used to compute the output activation of the LSTM unit. The following output function is used:
\begin{equation}
\Gamma _{o}^{t}=\sigma \left ( W_o\left [ h^{t-1},x^{t} \right ] +b_{o}\right ).
\end{equation}
The weight matrices and bias vector parameters are needed to be learned during training.
This paper updates the current working state as the following formula:
\begin{equation}\label{eq_hidden}
h^{t}=\Gamma _{o}^{t}\ast\tanh\left ( c^{t}\right ).
\end{equation}
The items stored in the current working state have an advantage in reading over those stored in long-term memory. In the time series modeling of scientific impact, the recent items stored in the short-term working state have an advantage over those stored in the long-term memory. The next step introduces the attention mechanism based on $h^{t}$.

\vspace{2ex}
\textbf{The Attention Model}.
The artificial attention mechanism, inspired by the attention behavior in neuroscience
, has been applied in deep learning for speech recognition, translation, and visual identification of object.
Broadly, attention mechanisms are components of prediction
systems that allow the system to focus on different subsets of the input sequentially. It aims to capture the critical points and focuses on the relevant parts more than the remote parts as a human does.
More specifically, the content-based attention generates attention distribution. Only part of a subset of the input information is focused. The attention function needs to be differentiable, so that everywhere of the input is focused, just to different extents.


The deep learning attention mechanism used in this paper works as follows: given an input $\vec{X}=\{x_{d}^{0},\cdots ,x_{d}^{t},\cdots, x_{d}^{T}\}$, the aforementioned LSTM units generate $\vec{h}=\{h_1, \cdots, h_t, \cdots, h_T\}$ to represent the hidden patterns of the input. The output is the summary of the $h_t$ focusing on information linked to the input. In this formulation, attention produces a fixed-length embedding of the input sequence by computing an adaptive weighted average of the state sequence $\vec{h}$.

The graphical representation of the attention model is shown in Fig.~\ref{subfig_attention}. The input $\vec{X}$ and the hidden layer $\vec{h}$ of LSTM network (a RNN composed of LSTM units) are the input of the attention model. Then, it computes the following formula:
\begin{equation}
a^{t}=\tanh\left ( W_a\left [ x^{t},h^{t} \right ]\right ),
\end{equation}
where $W_a$ is the weight matrix. An important remark here is that each $a^t$ is computed independently without looking at the other $x^{t'}$ for $t'\neq t$. Then, each $a^t$ is linked to a Softmax layer, which function is given by:
\begin{equation}
\alpha^t=\frac{e^{a^t}}{\sum_{t}e^{a^t}}, \text{for}~t=1,\cdots,T
\end{equation}
where $\sum_{t}\alpha^t=1$, the $\alpha^t$ is the softmax of the $a^t$ projected on a learned direction.
The output is a weighted arithmetic mean of the input, and the weights reflects the relevance of $\vec{h}$ and the input. It is calculated as the following formula:
\begin{equation}\label{eq_output}
O=\sum_{t}\alpha^t x_t.
\end{equation}
Finally, the popularity of item $d$ at time $t$ is given by the prediction $f(d|\vec{X}, t)=O$.

\begin{table*}[thbp]
\centering
\caption{The performance of various models on the data set.}
\label{table}
\begin{tabular}{|c|c|c|c|c|c|c|c|c|c|c|}
\hline
 & \multicolumn{2}{|c|}{$t=1$} & \multicolumn{2}{|c|}{$t=2$} & \multicolumn{2}{|c|}{$t=3$} & \multicolumn{2}{|c|}{$t=4$} & \multicolumn{2}{|c|}{$t=5$} \\
\hline
Models & MAPE & ACC & MAPE & ACC & MAPE & ACC & MAPE & ACC & MAPE & ACC \\
\hline
RPP & 0.219 & 0.819 & 0.381 & 0.661 & 0.686 & 0.524 & 0.904 & 0.433 & 1.376 & 0.370 \\
\hline
SVR & 0.195 & 0.814 & 0.252 & 0.664 & 0.296 & 0.579 & 0.331 & 0.528 & 0.362 & 0.493 \\
\hline
LR & 0.136 & 0.924 & 0.207 & 0.752 & 0.269 & 0.629 & 0.330 & 0.540 & 0.386 & 0.482 \\
\hline
CART & 0.131 & 0.913 & 0.202 & 0.758 & 0.256 & 0.634 & 0.297 & 0.549 & 0.328 & 0.489 \\
\hline
RNN & 0.123 & 0.940 & 0.185 & 0.804 & 0.234 & 0.703 & 0.298 & 0.590 & 0.317 & 0.551 \\
\hline
DLAM & \textbf{0.121} & \textbf{0.960} & \textbf{0.168} & \textbf{0.849} & \textbf{0.203} & \textbf{0.757} & \textbf{0.231} & \textbf{0.693} & \textbf{0.255} & \textbf{0.643} \\
\hline
\end{tabular}
\end{table*}

%
%

%


\subsection{Key Factor in Quantifying Long-term Impact}
As widely acknowledged, the citation distribution follows the power-law distribution. This finding leads the way of research in this domain. Researchers try to simulate the citation distribution as the power-law distribution.
This paper changes the line of thinking. Although the number of research papers has exploded, the reading time of scientists has not. The attention shifts toward the top $1$\% over time~\cite{barabasi2012publishing}.
Even though the citation distribution follows the power-law distribution, attention is also vital in quantifying the long-term scientific impact.

In the fact of limited attention, Matthew effect dominates in quantifying the long-term scientific impact. The experiments will confirm it. The citation count captures the inherent differences between papers, accounting for the perceived novelty and importance of a paper. The ``rich-get-richer" phenomenon summarizes the Matthew effect of accumulated advantage, i.e., previously accumulated attention triggers more subsequent attentions~\cite{Crane2008Robust}. It is in fact that the highly popular items are more visible and more likely to be viewed than others. The proposed model emphasizes highly cited papers under limited attention. The memory cell in the LSTM unit considers the long-term dependencies. As shown in Eq.~(\ref{eq_long}), previously accumulated attention stored in the long-term memory triggers more subsequent attention. What is more, the attention model, which focuses on the most popular part of the time series as Eq.~(\ref{eq_output}) does, also emphasizes the Matthew effect.

\section{Experiments}\label{sec:experiment}
This section demonstrates the effectiveness of putting particular emphasis on the vital factor in quantifying the long-term scientific impact.

\subsection{Dataset}
The authors extract the data from an academic search and mining platform called AMiner and construct a real large-scale scholarly dataset\footnote{https://www.aminer.cn/data}. The full graph of citation network contained in this dataset has about $2$ million vertices (papers) and $8$ million edges (citations). In detail, the dataset is composed of $2,092,356$ digitalized papers spanning from $1936$ to $2016$ (for more than $80$ years), and $8,024,869$ citations between them. By convention, the authors eliminate those papers with less than $5$ citations during the first $5$ years after publication and only retain the remaining papers as the training data. As a result, $143,902$ papers published in $1956$ to $2015$ are retained.

\subsection{Baseline Models and Evaluation Metrics}\label{evaluation}
To compare the predictive performance of the proposed attention model against other models, we introduce several published models that have been used to predict scientific impact. Specifically, the comparison methods in the experiments are LR, CART, SVR (the three basic machine learning methods used in \cite{Yan2011Citation}), RPP~\cite{wang2013quantifying,shen2014modeling}, and RNN \cite{Xiao2017Modeling}. The advantage of deep learning is the utilization of various features. For the sake of fairness, the authors only use the citation count records and the same feature used in~\cite{wang2013quantifying,shen2014modeling}.

This paper uses two basic metrics for scientific impact evaluation: Mean Absolute Percentage Error (MAPE) and Accuracy (ACC). Let $n_{d}^{t}$ be the observed citations of paper $d$ up to time $t$, and $\hat{n}_{d}^{t}$ be the predicted one. The MAPE measures the average deviation between the predicted and observed citations over all papers. For a dataset of $M$ papers, the MAPE is given by:
\begin{equation}
\mathrm{MAPE} =  \frac{1}{M}\sum_{d=1}^{M}\left | \frac{\hat{n}_{d}^{t}-n_{d}^{t}}{n_{d}^{t}} \right |.
\end{equation}

ACC measures the fraction of papers correctly predicted under a given error tolerance $\epsilon$. Specifically, the accuracy of citation prediction over $M$ papers is defined as:
\begin{equation}
\mathrm{ACC} =  \frac{1}{M}\sum_{d=1}^{M}\mathrm{I}\left [\left | \frac{\hat{n}_{d}^{t}-n_{d}^{t}}{n_{d}^{t}} \right |\leq \epsilon \right ],
\end{equation}
where $\mathrm{I}[\theta]$ is an indicator function which returns $1$ if the statement $\theta$ is true, otherwise returns $0$. We find that our method always outperforms regardless the value of $\epsilon$. In this paper, we set $\epsilon = 0.3$.

\subsection{Model Setting}
The experiment results show that the longer the duration of the training set, the better the long-term prediction performance. This paper sets the training period as $5$ years and then predict the citation counts for each paper from the $1^\mathrm{st}$ to $5^\mathrm{th}$ after the training period. For example, $t=1$ means that the first observation year after the training period. 
In the experiments, the features with positive contributions are the citation history, the h-index of the paper author, and the level of the publication journal. For the convenience of performance comparison, the input feature used here is the citation history for every sub-window of length $10$ years.
The value of the parameter $L$ is $2$. The loss function used here is MAPE. Adadelta is the gradient descent optimization algorithm. The attention layer is fully connected and uses tanh activation.

\begin{figure*}[htb]
\centering
\subfigure[ACC comparision.]{
\includegraphics[width=0.42\textwidth]{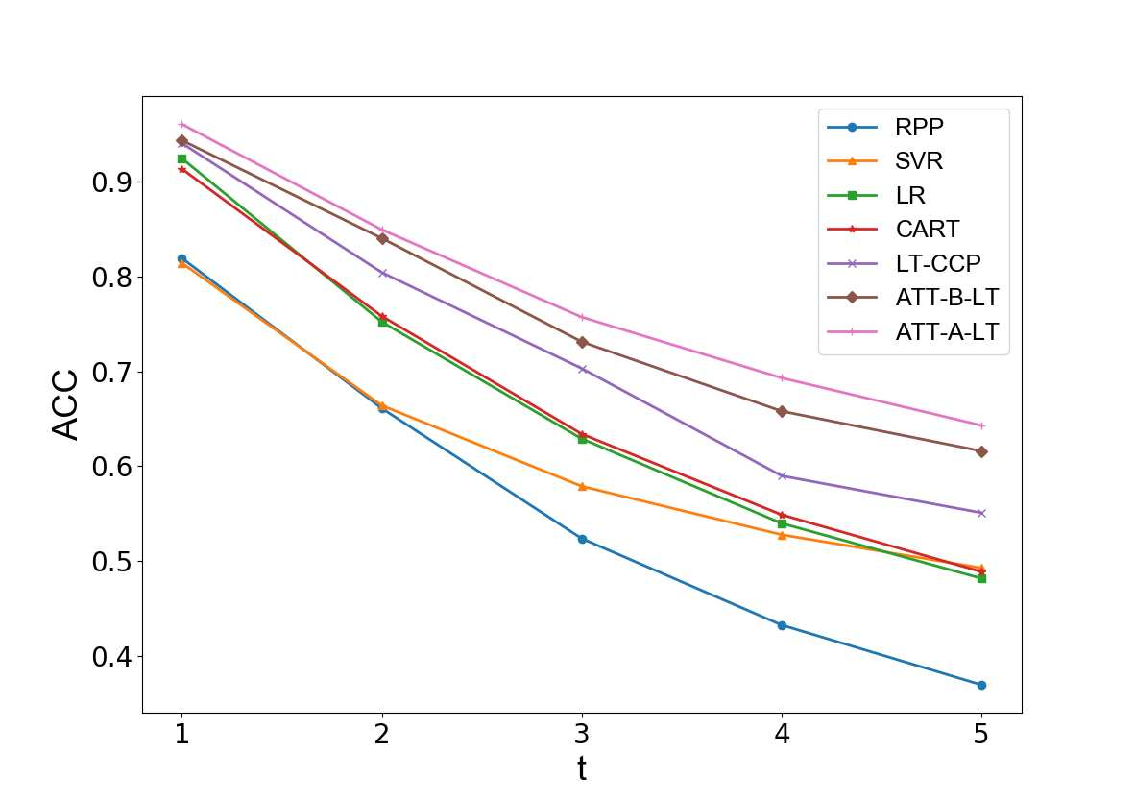}
\label{fig_acc}}
\subfigure[MAPE comparision.]{
\includegraphics[width=0.42\textwidth]{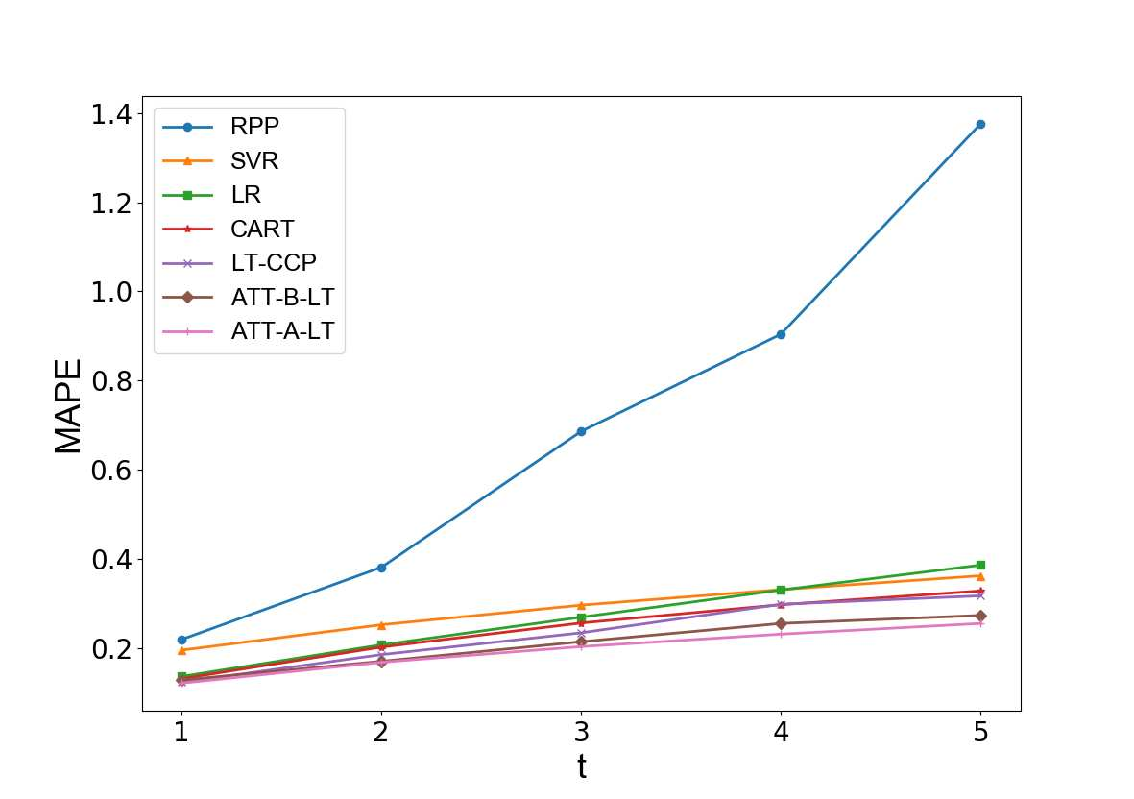}
\label{fig_mape}}
\subfigure[LT-CCP (RNN with LSTM).]{
\includegraphics[width=0.3\textwidth]{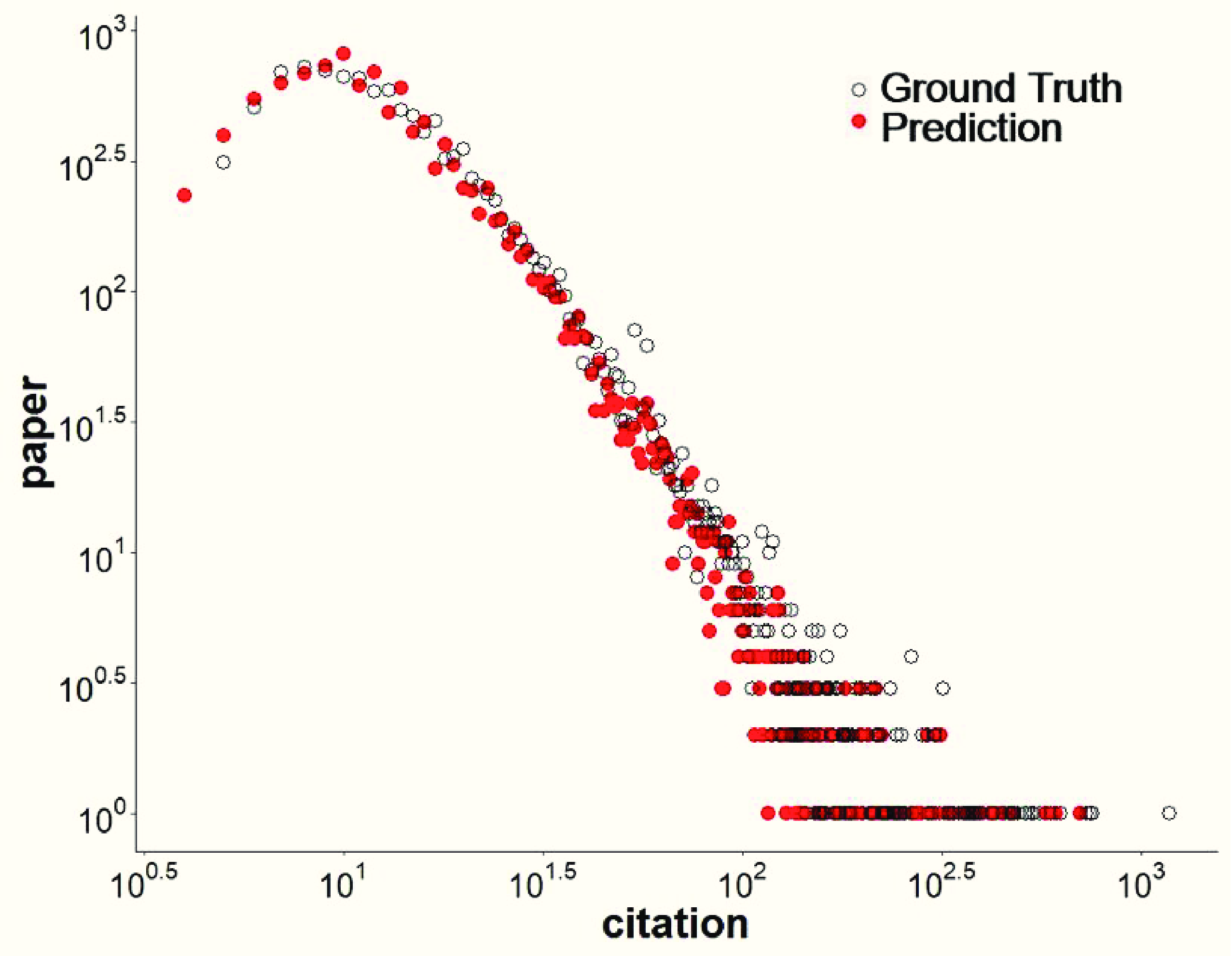}
\label{fig_a1}}
\subfigure[ATT-B-LT.]{
\includegraphics[width=0.3\textwidth]{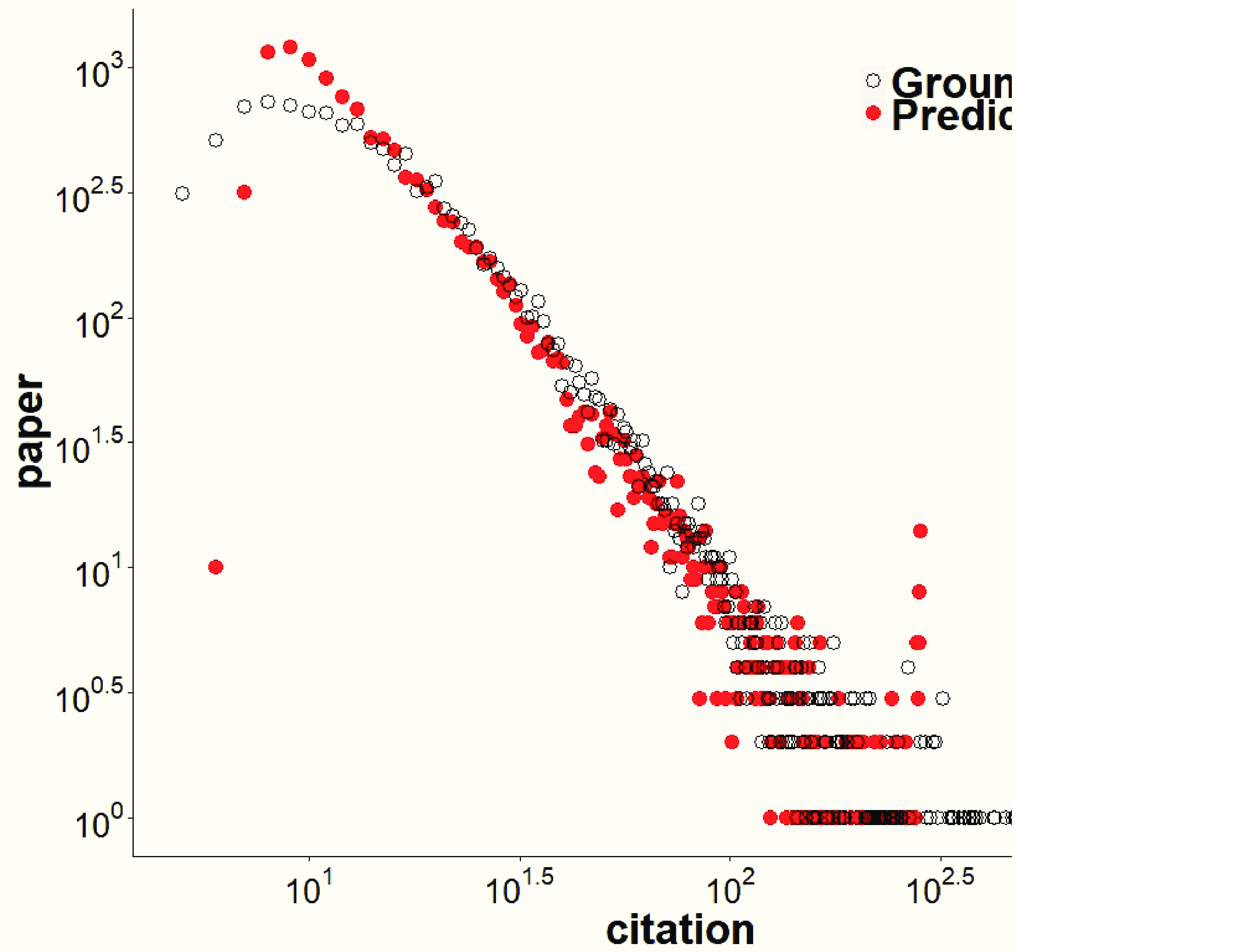}
\label{fig_a2}}
\subfigure[ATT-A-LT (DLAM).]{
\includegraphics[width=0.3\textwidth]{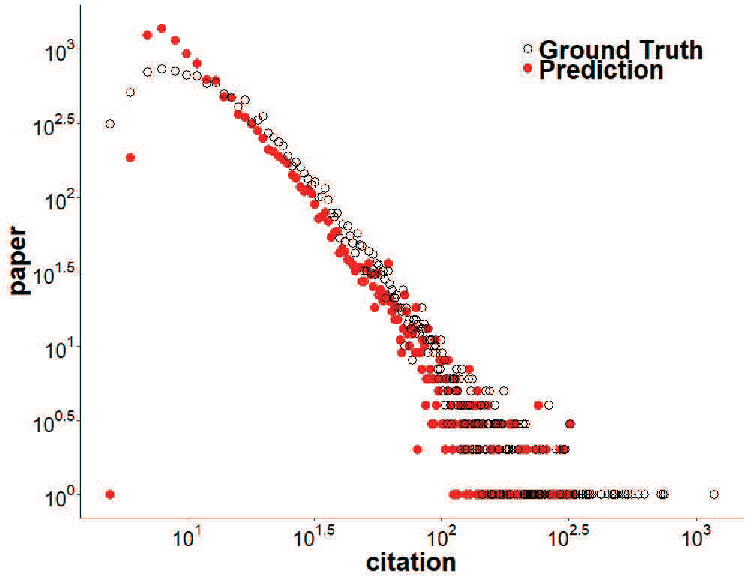}
\label{fig_a3}}
\caption{The performance comparison in citation count prediction.}
\label{fig_experiment2}
\end{figure*}

\subsection{Results}
As shown in the Table.~\ref{table}, the proposed model exhibits the best performance in terms of ACC in all the situations of $t=1$, $2$, $3$, $4$, and $5$. It means that the DLAM consistently achieves the higher accuracy than other models across different observation time. What is more, the proposed model also exhibits the best performance in terms of MAPE in all the situations mentioned above. That is, the proposed model achieves higher accuracy and lowers error rates simultaneously.
In the experiments, all the models used for comparison achieve acceptable low error rates, except RPP. RPP can avoid this problem with prior~\cite{shen2014modeling}, which incorporates conjugate prior for the fitness parameter. However, the RPP with prior does not improve the ACC performance. Overall, the proposed model also outperforms than RPP with prior.

Compared to the other methods in terms of ACC and MAPE, the proposed model increases with the number of years after the training period. Compare to RNN (the most efficient method certified in recent works), the proposed model achieves a few performance improvements about $1.65\%$ in terms of MAPE, and about $2.13\%$ in terms of ACC, when $t=1$. However, in the situation of $t=5$, the proposed model achieves significant performance improvement about $24.31\%$ in terms of MAPE, and about $16.7\%$ in terms of ACC. In other words, the proposed model shows much superiority than other models in scientific impact prediction, especially in the \textbf{long-term} situation.



\section{Further Exploration}
\vspace{1ex}
~~~~\textbf{Effectiveness of the attention mechanism.}
The authors remove the attention module of the proposed model to verify the effectiveness of the attention mechanism. The remainder is RNN with LSTM units (labeled as LT-CCP), which is proven to be useful in long-term citation count prediction.
In the next step, we add the attention mechanism in two different ways. Firstly, we add the attention module before the RNN module, which is labeled as ATT-B-LT (attention before LT-CCP). In a second way, we add the attention module after the RNN module, which is labeled as ATT-A-LT (attention after LT-CCP).
As shown in Fig.~\ref{fig_mape} and Fig.~\ref{fig_acc}, the ACC is increased, and the corresponding MAPE is decreased. Both ATT-B-LT and ATT-A-LT perform better than LT-CCP in terms of MAPE and ACC. Introducing the attention module improves the ability of scientific impact prediction. The effectiveness of the attention mechanism is verified.

In addition, we can see that the ATT-A-LT performs better than ATT-B-LT. It indicates that the deep learning model can learn the implicit features underlying the citation records, which provide a further boost in the performance.

\vspace{1ex}
\textbf{Analysis of the citation distribution.} We illustrate the actual and the predicted citations distribution of LT-CCP (RNN with LSTM), ATT-B-LT and ATT-A-LT (DLAM) when $t=5$ in Fig.~\ref{fig_a1}, Fig.~\ref{fig_a2} and Fig.~\ref{fig_a3} respectively.
The LT-CCP (RNN with LSTM) illustrated in Fig.~\ref{fig_a1} shows the best simulation of the power-law distribution. But the ATT-B-LT shown in Fig.~\ref{fig_a2} and the ATT-A-LT (DLAM) shown in Fig.~\ref{fig_a3} present bad simulation of the power-law distribution.
The results show that LT-CCP (RNN with LSTM) matches very well with that of real citations, but the ATT-B-LT and the ATT-A-LT (DLAM) don't.
Usually, it is believed that more similar to the power-law distribution, the whole result is more better. At first glance, it seems that LT-CCP (RNN with LSTM) performs the best.

However, the first thought is wrong. As verified in Fig.~\ref{fig_acc} and Fig.~\ref{fig_mape}, the LT-CCP (RNN with LSTM) performs the worst.
In fact, the LT-CCP only has better fitting effect on the papers with little citation counts. On the contrary, the ATT-B-LT and ATT-A-LT (DLAM) have better fitting effect on the highly cited papers. The methods with attention mechanism achieve better overall performance. It is more accordant with practical prediction requirements that a few papers occupy vast number of citations. It further proves the effectiveness of the attention model.
The experimental results indicate that we need to change the fixed pattern of thinking in quantifying long-term scientific impact.

\section{Conclusion}\label{conclusion}
Scientific impact evaluation is always a key point in decision making concerning with recruitment and funding in the scientific community. Based on big data based empirical analysis, science of science provides quantitative understanding of the scientific impact.
In this paper, the authors introduce the attention mechanism in long-term scientific impact prediction, and verify its effectiveness. More importantly, this paper provides us great insights in understanding the key factor in quantifying long-term scientific impact.
Usually, it is believed that the citation distribution is more similar to the power-law distribution, the whole result is more better. However, the experimental results in the paper discredit this conclusion. In the future research work, we need to change the fixed pattern of thinking in quantifying long-term scientific impact, and make better use of limited attention to better stand on the shoulders of giants.

\section*{Acknowledgments}
The work is supported by the National Natural Science Foundation of China (NSFC) under Grant No.~$61806111$ and No.~$61806109$, and NSFC for Distinguished Young Scholar under Grant No.~$61825602$.

\bibliographystyle{ACM-Reference-Format}
\bibliography{sample-base}


\begin{thebibliography}{27}


\ifx \showCODEN    \undefined \def \showCODEN     #1{\unskip}     \fi
\ifx \showDOI      \undefined \def \showDOI       #1{#1}\fi
\ifx \showISBNx    \undefined \def \showISBNx     #1{\unskip}     \fi
\ifx \showISBNxiii \undefined \def \showISBNxiii  #1{\unskip}     \fi
\ifx \showISSN     \undefined \def \showISSN      #1{\unskip}     \fi
\ifx \showLCCN     \undefined \def \showLCCN      #1{\unskip}     \fi
\ifx \shownote     \undefined \def \shownote      #1{#1}          \fi
\ifx \showarticletitle \undefined \def \showarticletitle #1{#1}   \fi
\ifx \showURL      \undefined \def \showURL       {\relax}        \fi
\providecommand\bibfield[2]{#2}
\providecommand\bibinfo[2]{#2}
\providecommand\natexlab[1]{#1}
\providecommand\showeprint[2][]{arXiv:#2}

\bibitem[\protect\citeauthoryear{Acuna, Allesina, and Kording}{Acuna
  et~al\mbox{.}}{2012}]%
        {acuna2012future}
\bibfield{author}{\bibinfo{person}{Daniel~E Acuna}, \bibinfo{person}{Stefano
  Allesina}, {and} \bibinfo{person}{Konrad~P Kording}.}
  \bibinfo{year}{2012}\natexlab{}.
\newblock \showarticletitle{Future impact: Predicting scientific success}.
\newblock \bibinfo{journal}{\emph{Nature}} \bibinfo{volume}{489},
  \bibinfo{number}{7415} (\bibinfo{year}{2012}), \bibinfo{pages}{201}.
\newblock


\bibitem[\protect\citeauthoryear{Bao, Shen, Huang, and Cheng}{Bao
  et~al\mbox{.}}{2013}]%
        {Bao2013Popularity}
\bibfield{author}{\bibinfo{person}{Peng Bao}, \bibinfo{person}{Hua~Wei Shen},
  \bibinfo{person}{Junming Huang}, {and} \bibinfo{person}{Xue~Qi Cheng}.}
  \bibinfo{year}{2013}\natexlab{}.
\newblock \showarticletitle{Popularity prediction in microblogging network: a
  case study on sina weibo}. In \bibinfo{booktitle}{\emph{International
  Conference on World Wide Web}}. \bibinfo{pages}{177--178}.
\newblock


\bibitem[\protect\citeauthoryear{Barab{\'a}si, Song, and Wang}{Barab{\'a}si
  et~al\mbox{.}}{2012}]%
        {barabasi2012publishing}
\bibfield{author}{\bibinfo{person}{Albert-L{\'a}szl{\'o} Barab{\'a}si},
  \bibinfo{person}{Chaoming Song}, {and} \bibinfo{person}{Dashun Wang}.}
  \bibinfo{year}{2012}\natexlab{}.
\newblock \showarticletitle{Publishing: Handful of papers dominates citation}.
\newblock \bibinfo{journal}{\emph{Nature}} \bibinfo{volume}{491},
  \bibinfo{number}{7422} (\bibinfo{year}{2012}), \bibinfo{pages}{40}.
\newblock


\bibitem[\protect\citeauthoryear{Bengio, Simard, and Frasconi}{Bengio
  et~al\mbox{.}}{1994}]%
        {bengio1994learning}
\bibfield{author}{\bibinfo{person}{Yoshua Bengio}, \bibinfo{person}{Patrice
  Simard}, {and} \bibinfo{person}{Paolo Frasconi}.}
  \bibinfo{year}{1994}\natexlab{}.
\newblock \showarticletitle{Learning long-term dependencies with gradient
  descent is difficult}.
\newblock \bibinfo{journal}{\emph{IEEE transactions on neural networks}}
  \bibinfo{volume}{5}, \bibinfo{number}{2} (\bibinfo{year}{1994}),
  \bibinfo{pages}{157--166}.
\newblock


\bibitem[\protect\citeauthoryear{Crane and Sornette}{Crane and
  Sornette}{2008}]%
        {Crane2008Robust}
\bibfield{author}{\bibinfo{person}{R Crane} {and} \bibinfo{person}{D
  Sornette}.} \bibinfo{year}{2008}\natexlab{}.
\newblock \showarticletitle{Robust dynamic classes revealed by measuring the
  response function of a social system}.
\newblock \bibinfo{journal}{\emph{Proceedings of the National Academy of
  Sciences of the United States of America}} \bibinfo{volume}{105},
  \bibinfo{number}{41} (\bibinfo{year}{2008}), \bibinfo{pages}{15649--53}.
\newblock


\bibitem[\protect\citeauthoryear{Dong, Johnson, and Chawla}{Dong
  et~al\mbox{.}}{2015}]%
        {dong2015will}
\bibfield{author}{\bibinfo{person}{Yuxiao Dong}, \bibinfo{person}{Reid~A
  Johnson}, {and} \bibinfo{person}{Nitesh~V Chawla}.}
  \bibinfo{year}{2015}\natexlab{}.
\newblock \showarticletitle{Will this paper increase your h-index?: Scientific
  impact prediction}. In \bibinfo{booktitle}{\emph{Proceedings of the eighth
  ACM international conference on web search and data mining}}. ACM,
  \bibinfo{pages}{149--158}.
\newblock


\bibitem[\protect\citeauthoryear{Fortunato, Bergstrom, B{\"o}rner, Evans,
  Helbing, Milojevi{\'c}, et~al\mbox{.}}{Fortunato et~al\mbox{.}}{2018}]%
        {fortunato2018science}
\bibfield{author}{\bibinfo{person}{Santo Fortunato}, \bibinfo{person}{Carl~T
  Bergstrom}, \bibinfo{person}{Katy B{\"o}rner}, \bibinfo{person}{James~A
  Evans}, \bibinfo{person}{Dirk Helbing}, \bibinfo{person}{Sta{\v{s}}a
  Milojevi{\'c}}, {et~al\mbox{.}}} \bibinfo{year}{2018}\natexlab{}.
\newblock \showarticletitle{Science of science}.
\newblock \bibinfo{journal}{\emph{Science}} \bibinfo{volume}{359},
  \bibinfo{number}{6379} (\bibinfo{year}{2018}).
\newblock


\bibitem[\protect\citeauthoryear{Garfield}{Garfield}{2001}]%
        {garfield2001impact}
\bibfield{author}{\bibinfo{person}{Eugene Garfield}.}
  \bibinfo{year}{2001}\natexlab{}.
\newblock \showarticletitle{Impact factors, and why they won't go away}.
\newblock \bibinfo{journal}{\emph{Nature}} \bibinfo{volume}{411},
  \bibinfo{number}{6837} (\bibinfo{year}{2001}), \bibinfo{pages}{522}.
\newblock


\bibitem[\protect\citeauthoryear{Hirsch}{Hirsch}{2005}]%
        {Hirsch2005An}
\bibfield{author}{\bibinfo{person}{J.~E. Hirsch}.}
  \bibinfo{year}{2005}\natexlab{}.
\newblock \showarticletitle{An Index to Quantify an Individual's Scientific
  Research Output}.
\newblock \bibinfo{journal}{\emph{Proceedings of the National Academy of
  Sciences of the United States of America}} \bibinfo{volume}{102},
  \bibinfo{number}{46} (\bibinfo{year}{2005}), \bibinfo{pages}{16569--16572}.
\newblock


\bibitem[\protect\citeauthoryear{Hunter, Smyth, Vu, and Asuncion}{Hunter
  et~al\mbox{.}}{2011}]%
        {hunter2011dynamic}
\bibfield{author}{\bibinfo{person}{David Hunter}, \bibinfo{person}{Padhraic
  Smyth}, \bibinfo{person}{Duy~Q Vu}, {and} \bibinfo{person}{Arthur~U
  Asuncion}.} \bibinfo{year}{2011}\natexlab{}.
\newblock \showarticletitle{Dynamic egocentric models for citation networks}.
  In \bibinfo{booktitle}{\emph{Proceedings of the 28th International Conference
  on Machine Learning (ICML-11)}}. \bibinfo{pages}{857--864}.
\newblock


\bibitem[\protect\citeauthoryear{Kuhn, Perc, and Helbing}{Kuhn
  et~al\mbox{.}}{2014}]%
        {kuhn2014inheritance}
\bibfield{author}{\bibinfo{person}{Tobias Kuhn}, \bibinfo{person}{Matja{\v{z}}
  Perc}, {and} \bibinfo{person}{Dirk Helbing}.}
  \bibinfo{year}{2014}\natexlab{}.
\newblock \showarticletitle{Inheritance patterns in citation networks reveal
  scientific memes}.
\newblock \bibinfo{journal}{\emph{Physical Review X}} \bibinfo{volume}{4},
  \bibinfo{number}{4} (\bibinfo{year}{2014}), \bibinfo{pages}{041036}.
\newblock


\bibitem[\protect\citeauthoryear{Larivi{\`e}re, Haustein, and
  B{\"o}rner}{Larivi{\`e}re et~al\mbox{.}}{2015}]%
        {lariviere2015long}
\bibfield{author}{\bibinfo{person}{Vincent Larivi{\`e}re},
  \bibinfo{person}{Stefanie Haustein}, {and} \bibinfo{person}{Katy
  B{\"o}rner}.} \bibinfo{year}{2015}\natexlab{}.
\newblock \showarticletitle{Long-distance interdisciplinarity leads to higher
  scientific impact}.
\newblock \bibinfo{journal}{\emph{Plos one}} \bibinfo{volume}{10},
  \bibinfo{number}{3} (\bibinfo{year}{2015}).
\newblock


\bibitem[\protect\citeauthoryear{Li, He, Tian, Chen, Qin, Wang, and Liu}{Li
  et~al\mbox{.}}{2018}]%
        {li2018towards}
\bibfield{author}{\bibinfo{person}{Zhuohan Li}, \bibinfo{person}{Di He},
  \bibinfo{person}{Fei Tian}, \bibinfo{person}{Wei Chen}, \bibinfo{person}{Tao
  Qin}, \bibinfo{person}{Liwei Wang}, {and} \bibinfo{person}{Tie-Yan Liu}.}
  \bibinfo{year}{2018}\natexlab{}.
\newblock \showarticletitle{Towards Binary-Valued Gates for Robust LSTM
  Training}. In \bibinfo{booktitle}{\emph{International Conference on Machine
  Learning}}. \bibinfo{pages}{3001--3010}.
\newblock


\bibitem[\protect\citeauthoryear{Matsubara, Sakurai, Prakash, Li, and
  Faloutsos}{Matsubara et~al\mbox{.}}{2012}]%
        {Matsubara2012Rise}
\bibfield{author}{\bibinfo{person}{Yasuko Matsubara}, \bibinfo{person}{Yasushi
  Sakurai}, \bibinfo{person}{B.~Aditya Prakash}, \bibinfo{person}{Lei Li},
  {and} \bibinfo{person}{Christos Faloutsos}.} \bibinfo{year}{2012}\natexlab{}.
\newblock \showarticletitle{Rise and fall patterns of information diffusion:
  model and implications}. In \bibinfo{booktitle}{\emph{ACM SIGKDD
  International Conference on Knowledge Discovery and Data Mining}}.
  \bibinfo{pages}{6--14}.
\newblock


\bibitem[\protect\citeauthoryear{Pan, Kaski, and Fortunato}{Pan
  et~al\mbox{.}}{2012}]%
        {pan2012world}
\bibfield{author}{\bibinfo{person}{Raj~Kumar Pan}, \bibinfo{person}{Kimmo
  Kaski}, {and} \bibinfo{person}{Santo Fortunato}.}
  \bibinfo{year}{2012}\natexlab{}.
\newblock \showarticletitle{World citation and collaboration networks:
  uncovering the role of geography in science}.
\newblock \bibinfo{journal}{\emph{Scientific reports}}  \bibinfo{volume}{2}
  (\bibinfo{year}{2012}), \bibinfo{pages}{902}.
\newblock


\bibitem[\protect\citeauthoryear{Peng}{Peng}{2016}]%
        {Bao2016Modeling}
\bibfield{author}{\bibinfo{person}{Bao Peng}.} \bibinfo{year}{2016}\natexlab{}.
\newblock \showarticletitle{Modeling and Predicting Popularity Dynamics via an
  Influence-based Self-Excited Hawkes Process}. In
  \bibinfo{booktitle}{\emph{ACM International on Conference on Information and
  Knowledge Management}}. \bibinfo{pages}{1897--1900}.
\newblock


\bibitem[\protect\citeauthoryear{Radicchi, Fortunato, and Castellano}{Radicchi
  et~al\mbox{.}}{2008}]%
        {radicchi2008universality}
\bibfield{author}{\bibinfo{person}{Filippo Radicchi}, \bibinfo{person}{Santo
  Fortunato}, {and} \bibinfo{person}{Claudio Castellano}.}
  \bibinfo{year}{2008}\natexlab{}.
\newblock \showarticletitle{Universality of citation distributions: Toward an
  objective measure of scientific impact}.
\newblock \bibinfo{journal}{\emph{Proceedings of the National Academy of
  Sciences}} \bibinfo{volume}{105}, \bibinfo{number}{45}
  (\bibinfo{year}{2008}), \bibinfo{pages}{17268--17272}.
\newblock


\bibitem[\protect\citeauthoryear{Shen, Wang, Song, and Barab{\'a}si}{Shen
  et~al\mbox{.}}{2014}]%
        {shen2014modeling}
\bibfield{author}{\bibinfo{person}{Huawei Shen}, \bibinfo{person}{Dashun Wang},
  \bibinfo{person}{Chaoming Song}, {and} \bibinfo{person}{Albert-L{\'a}szl{\'o}
  Barab{\'a}si}.} \bibinfo{year}{2014}\natexlab{}.
\newblock \showarticletitle{Modeling and predicting popularity dynamics via
  reinforced poisson processes}. In \bibinfo{booktitle}{\emph{Twenty-eighth
  AAAI conference on artificial intelligence}}.
\newblock


\bibitem[\protect\citeauthoryear{Sutskever, Vinyals, and Le}{Sutskever
  et~al\mbox{.}}{2014}]%
        {sutskever2014sequence}
\bibfield{author}{\bibinfo{person}{Ilya Sutskever}, \bibinfo{person}{Oriol
  Vinyals}, {and} \bibinfo{person}{Quoc~V Le}.}
  \bibinfo{year}{2014}\natexlab{}.
\newblock \showarticletitle{Sequence to sequence learning with neural
  networks}. In \bibinfo{booktitle}{\emph{Advances in Neural Information
  Processing Systems}}. \bibinfo{pages}{3104--3112}.
\newblock


\bibitem[\protect\citeauthoryear{Szabo and Huberman}{Szabo and
  Huberman}{2010}]%
        {Szabo2010Predicting}
\bibfield{author}{\bibinfo{person}{Gabor Szabo} {and}
  \bibinfo{person}{Bernardo~A. Huberman}.} \bibinfo{year}{2010}\natexlab{}.
\newblock \bibinfo{booktitle}{\emph{Predicting the popularity of online
  content}}. Vol.~\bibinfo{volume}{53}.
\newblock \bibinfo{publisher}{Communications of the ACM}. 80--88 pages.
\newblock


\bibitem[\protect\citeauthoryear{Tang, Zhang, Yao, Li, Zhang, and Su}{Tang
  et~al\mbox{.}}{2008}]%
        {Tang2008ArnetMiner}
\bibfield{author}{\bibinfo{person}{Jie Tang}, \bibinfo{person}{Jing Zhang},
  \bibinfo{person}{Limin Yao}, \bibinfo{person}{Juanzi Li}, \bibinfo{person}{Li
  Zhang}, {and} \bibinfo{person}{Zhong Su}.} \bibinfo{year}{2008}\natexlab{}.
\newblock \showarticletitle{ArnetMiner:extraction and mining of academic social
  networks}. In \bibinfo{booktitle}{\emph{ACM SIGKDD International Conference
  on Knowledge Discovery and Data Mining}}. \bibinfo{pages}{990--998}.
\newblock


\bibitem[\protect\citeauthoryear{Vu, Asuncion, Hunter, and Smyth}{Vu
  et~al\mbox{.}}{2011}]%
        {Vu2011Dynamic}
\bibfield{author}{\bibinfo{person}{Duy~Q Vu}, \bibinfo{person}{Arthur~U
  Asuncion}, \bibinfo{person}{David~R Hunter}, {and} \bibinfo{person}{Padhraic
  Smyth}.} \bibinfo{year}{2011}\natexlab{}.
\newblock \showarticletitle{Dynamic egocentric models for citation networks}.
  In \bibinfo{booktitle}{\emph{International Conference on International
  Conference on Machine Learning}}. \bibinfo{pages}{857--864}.
\newblock


\bibitem[\protect\citeauthoryear{Wang, Song, and Barab{\'a}si}{Wang
  et~al\mbox{.}}{2013}]%
        {wang2013quantifying}
\bibfield{author}{\bibinfo{person}{Dashun Wang}, \bibinfo{person}{Chaoming
  Song}, {and} \bibinfo{person}{Albert-L{\'a}szl{\'o} Barab{\'a}si}.}
  \bibinfo{year}{2013}\natexlab{}.
\newblock \showarticletitle{Quantifying long-term scientific impact}.
\newblock \bibinfo{journal}{\emph{Science}} \bibinfo{volume}{342},
  \bibinfo{number}{6154} (\bibinfo{year}{2013}), \bibinfo{pages}{127--132}.
\newblock


\bibitem[\protect\citeauthoryear{Xiao, Yan, Li, and Jin}{Xiao
  et~al\mbox{.}}{2016}]%
        {Xiao2016}
\bibfield{author}{\bibinfo{person}{Shuai Xiao}, \bibinfo{person}{Junchi Yan},
  \bibinfo{person}{Changsheng Li}, {and} \bibinfo{person}{Bo Jin}.}
  \bibinfo{year}{2016}\natexlab{}.
\newblock \showarticletitle{On Modeling and Predicting individual Paper
  Citation Count over Time}. In \bibinfo{booktitle}{\emph{Twenty-Fifth
  International Joint Conference on Artificial Intelligence (IJCAI-16)}}.
  \bibinfo{pages}{2676--2682}.
\newblock


\bibitem[\protect\citeauthoryear{Xiao, Yan, Yang, Zha, and Chu}{Xiao
  et~al\mbox{.}}{2017}]%
        {Xiao2017Modeling}
\bibfield{author}{\bibinfo{person}{Shuai Xiao}, \bibinfo{person}{Junchi Yan},
  \bibinfo{person}{Xiaokang Yang}, \bibinfo{person}{Hongyuan Zha}, {and}
  \bibinfo{person}{Stephen~M Chu}.} \bibinfo{year}{2017}\natexlab{}.
\newblock \showarticletitle{Modeling the Intensity Function of Point Process
  Via Recurrent Neural Networks.}. In \bibinfo{booktitle}{\emph{Thirty-First
  AAAI Conference on Artificial Intelligence}}, Vol.~\bibinfo{volume}{17}.
  \bibinfo{pages}{1597--1603}.
\newblock


\bibitem[\protect\citeauthoryear{Yan, Tang, Liu, Shan, and Li}{Yan
  et~al\mbox{.}}{2011}]%
        {Yan2011Citation}
\bibfield{author}{\bibinfo{person}{Rui Yan}, \bibinfo{person}{Jie Tang},
  \bibinfo{person}{Xiaobing Liu}, \bibinfo{person}{Dongdong Shan}, {and}
  \bibinfo{person}{Xiaoming Li}.} \bibinfo{year}{2011}\natexlab{}.
\newblock \showarticletitle{Citation count prediction: learning to estimate
  future citations for literature}. In \bibinfo{booktitle}{\emph{ACM
  International Conference on Information and Knowledge Management}}.
  \bibinfo{pages}{1247--1252}.
\newblock


\bibitem[\protect\citeauthoryear{Yegros-Yegros, Rafols, and
  D'Este}{Yegros-Yegros et~al\mbox{.}}{2015}]%
        {yegros2015does}
\bibfield{author}{\bibinfo{person}{Alfredo Yegros-Yegros},
  \bibinfo{person}{Ismael Rafols}, {and} \bibinfo{person}{Pablo D'Este}.}
  \bibinfo{year}{2015}\natexlab{}.
\newblock \showarticletitle{Does interdisciplinary research lead to higher
  citation impact? The different effect of proximal and distal
  interdisciplinarity}.
\newblock \bibinfo{journal}{\emph{PloS one}} \bibinfo{volume}{10},
  \bibinfo{number}{8} (\bibinfo{year}{2015}).
\newblock


\end{thebibliography}

\end{document}